\documentclass[prl,twocolumn,graphicx,amssymb,floatfix]{revtex4}

\begin{document}
{\bf  Comment on ``Weak value amplification is suboptimal for estimation and detection'' }

In a recent Letter, Ferrie and Combes \cite{fer} compared a statistical analysis based on weak values with the more standard method of maximum-likelihood estimation and found the latter better. They argued that ``postselection in general can only hinder the ability to perform the task of either detect or estimate". The derivation included an explicit form for the weak value, Eq (2), but the authors were ``assuming it is real". This assumption was not explained or justified in any way, even though it is evident from Eq (2) that the weak value is a complex number. 

The scenario of a complex, or even pure imaginary, weak value is rather different than one where it is real \cite{AV90,joz}. The distinctions has led to a number of proposals for new experimental schemes \cite{brunner,phase,strubi} and the particular advantages of imaginary weak values were widely studied \cite{dressel,snr,Nishizawa,jordan}. More importantly, most (if not all) of the experiment showing enhanced precision based of weak values were done with imaginary weak values \cite{e1,e2,e3,e4,e5,e6,e7,e8,e9}. 

The assumption of a real weak value led to an erroneous description of the typical analysis in a weak measurement experiment: ``The WVA approach is to take all the data $(r, f)$ and consider the distribution of the meter variable conditioned on the outcomes of the $A$ system:$ Pr(r|f,x)$." For imaginary weak values a different variable is observed. Usually it is the conjugated variable to $q$, which is related to $r$ via Eq. (3) in the letter. So Eq. (1) cannot be used to analyze the results of such scenarios.

Ignoring the conjugate variable can also lead to some confusion regarding the uncertainty of the meter. For a Gaussian wave function with variance $\sigma^2$ in $q$, the uncertainty in the conjugate variable is $\sigma^{-1}$. So stating that ``weak measurement (without
postselection and anomalous weak values) can mitigate the effect of technical noise for large enough $\sigma^2$" might not assist an experimentalist struggling to minimize uncertainties in a setup.

It is true that the imaginary part is not directly related to rare postselection and large weak values, even though these phenomena might be connected by practical consideration \cite{e3}. Thus it can be argued that it is a different effect than the ``weak value amplification". However, the imaginary part is directly related to postselection.  The authors consider postselection as ``a necessary ingredient for weak value amplification" but neglected to observe that it is also required for imaginary weak values. Thus, their comments that ``Postselection cannot aid in detect or estimate for any interaction parameter" are unsupported, to say the least. Furthermore, their result is irrelevant to any scenario where the weak values are complex, which are much more widespread than a pure real weak value.

Y. Kedem\\
Nordic Institute for Theoretical Physics (NORDITA)\\
Roslagstullsbacken 23, S-106 91 Stockholm, Sweden\\

\end{document}